\documentclass[fleqn,usenatbib]{mnras}

\usepackage{newtxtext,newtxmath}

\usepackage[T1]{fontenc}

\DeclareRobustCommand{\VAN}[3]{#2}
\let\VANthebibliography\thebibliography
\def\thebibliography{\DeclareRobustCommand{\VAN}[3]{##3}\VANthebibliography}

\usepackage{graphicx}	
\usepackage{amsmath}	
\usepackage{amssymb}	
\usepackage{ae,aecompl}
\usepackage{gensymb}
\usepackage{longtable}
\usepackage{graphbox}

\title[Cygnus Loop Distance]{An Updated Distance to the Cygnus Loop based on Gaia Early DR3}

\author[R. A. Fesen et al.]{
Robert A. Fesen,$^{1}$\thanks{E-mail: robert.fesen@dartmouth.edu}
Kathryn E.\ Weil,$^{2}$
Ignacio Cisneros,$^{1}$
William P.\ Blair,$^{3}$
and John C.\ Raymond,$^{4}$
\\
$^{1}$Department of Physics and Astronomy, 6127 Wilder Laboratory, Dartmouth College, Hanover, NH 03755, USA \\
$^{2}$Department of Physics and Astronomy, Purdue University, 525 Northwestern Avenue, West Lafayette, IN 47907 USA \\
$^{3}$The Henry A. Rowland Department of Physics \& Astronomy,  Johns Hopkins University, 3400 N. Charles Street, Baltimore, MD, 21218, USA \\
$^{4}$ Harvard-Smithsonian Center for Astrophysics, 60 Garden St., Cambridge, MA 02138, USA \\
}

\date{Accepted 2021 July 13. Received 2021 July 11; in original form 2021 June 26}

\pubyear{2015}

\begin{document}
\label{firstpage}
\pagerange{\pageref{firstpage}--\pageref{lastpage}}
\maketitle

\begin{abstract}
We present a revised distance to the Cygnus Loop supernova remnant of
$725\pm15$ pc based on Gaia Early Data Release 3 parallax measurements (EDR3) for
several stars previously found to be located either inside or behind the
supernova based on the presence of high-velocity absorption lines in their
spectra. This revised distance estimate and error means the Cygnus Loop remnant
now has an estimated distance uncertainty comparable to that of its $\simeq$18
pc radius.
\end{abstract}




\section{Introduction}

The Galactic supernova remnant (SNR) G74.0-8.5,  commonly known as the Cygnus
Loop or Veil Nebula, is thought to be a middle-age remnant with an estimated
age $\approx2 \times 10^{4}$ yr. The distance to the Cygnus Loop is a key
parameter in estimating its mass, energy and evolutionary state, as well as its
shock velocities determined from filament proper motions.  The most widely
adopted value for the Cygnus Loop's distance had long been 770 pc based on a
kinematic investigation by \citet{Minkowski1958}. However, subsequent distance
estimates have ranged between 500 to 1000 pc
\citep{Blair2009,Salvesen2009,Medina2014,Raymond2015}. See \citet{Fesen2018a}
for a full review.

The most recent distance estimate is based on
Na I 5890, 5896 \AA \ and Ca II 3934 \AA \  absorption
lines with velocities ranging from $-150$ to $+240$ km s$^{-1}$ in the spectra
of stars with projected locations toward the remnant \citep{Fesen2018b}. 
The presence of both red-
and blue-shifted absorptions means the stars must be located behind the Cygnus
Loop's expanding front and rear hemispheres. Combining Gaia Data Release 2
(DR2) parallax measurements \citep{Gaia2018} for three such stars plus the
estimated distance for a B7 star BD+31~4224 located along the remnant's
northwestern limb  \citep{Fesen2018a}, led to an estimated a distance to the
Cygnus Loop's center of $735 \pm 25$ pc and a model where the remnant's eastern
side was tilted toward us. However, this result involved
simple parallax inversion and not a Bayesian analysis \citep{Luri2018}.

Here we present an updated Cygnus Loop distance using Gaia Early Data Release 3
(EDR3; \citealt{Gaia2021}) of these same stars with parallaxes less than half the
uncertainty of DR2 parallaxes for stars with G magnitudes between 9 and 12 \citep{Lindegren2021}.
We adopt the star naming convention of \citet{Fesen2018b}
along with their conclusions about the locations of these stars with respect to
the Cygnus Loop remnant.

\begin{table*}
   \caption{Stars Exhibiting High-Velocity Absorption Features Associated with the Cygnus Loop Supernova Remnant}
\label{tab:table1}
\begin{tabular}{lcccccccc}
 Star      &   V     & Gaia EDR3 & RA      & Dec     & EDR3 & \multicolumn{2}{c}{\underline{~~~~~~~~~Distance (pc)~~~~~~~} } & Location \\
 ID$^{a}$  &  (mag)  & ID    & (J2000) & (J2000) & Parallax (mas) & Nominal & Modelled$^{b}$ & Relative to SNR \\
 \hline
 BD+31 4224  & 9.6  & 1859945414417986432 & 20:47:51.817  & +32:14:11.33  & $1.3569\pm 0.0251$  & $737^{+14}_{-13}$ & $726^{+13}_{-11}$  & at or inside  \\
 Star X      & 9.5  & 1858767429455484288 & 20:56:44.629  & +30:41:14.33  & $1.3478\pm 0.0209$  & $742^{+12}_{-11}$ & $732^{+12}_{-11}$  &  just behind  \\
 Star Y      & 11.3 & 1864900642756024832 & 20:55:51.948  & +31:43:27.40  & $1.3592\pm 0.0234$  & $736^{+13}_{-12}$ & $708^{+10}_{-10}$  &  just behind  \\
 Star Z      & 10.7 & 1865692531649104896 & 20:52:27.557  & +31:56:29.48  & $1.1749\pm 0.0244$  & $851^{+18}_{-17}$ & $835^{+20}_{-17}$  & far behind   \\
 KPD 2055+3111& 14.1& 1864830136570923904 & 20:57:26.889  & +31:22:52.56  & $1.2121\pm 0.0299$  & $825^{+21}_{-20}$ & $819^{+21}_{-18}$  &  far behind  \\
 \end{tabular}
\begin{flushleft}
$^{a}${References: BD+31 4224: \citet{Fesen2018a};  X,Y,Z: \citet{Fesen2018b}; KPD 2055 +3111: \citet{Blair2009} } \\
$^{b}${Direction-dependent distances from \citet{Bailer2021}.}
 \end{flushleft}
 \end{table*}

\section{Gaia Early DR3 Parallax Measurements}

Table 1 lists Gaia EDR3 parallaxes and parallax errors for three stars, dubbed
X, Y, and Z, with projected locations along the eastern half of the Cygnus Loop
that \citet{Fesen2018b} found to exhibit high-velocity \ion{Na}{1} and
\ion{Ca}{2} absorption components. The table also includes Gaia parallaxes for
the sdO star KPD 2055+3111 found by \citet{Blair2009} which also exhibited
high-velocity absorption features attributed to the Cygnus Loop, and the B7 V
star BD+31~4224 proposed by \citet{Fesen2018a} to physically lie inside or
physical contact with the remnant's shocks along the northwestern
limb. 
Coordinates and V magnitudes for these five stars are listed along
with the nominal geometric parallax distances (d = 1/p) and corrected distance
estimates following a probabilistic analysis as described by \citet{Bailer2015}
and tabulated for EDR3 by \citet{Bailer2021} using a three-dimensional model of
the Galaxy and a direction-dependent prior of distance.

\section{Discussion}

Since the star KPD 2055+3111 and Star Z lie at corrected EDR3 of distances of
819 and 835 pc, respectively, they must lie well behind the the Cygnus Loop
remnant based on the appearance of SNR related high-velocity absorption lines,
similar to those seen in the much closer stars, X and Y. Hence, they are not
particularly helpful in constraining the remnant's distance except in setting a
firm maximum distance of $\simeq800$ pc. We note that a previous search for
high-velocity absorption components associated with the remnant failed for
stars with distances less than $\sim$600 pc \citep{Welsh2002} which sets a
rough lower limit.

Adopting the situation described by \citet{Fesen2018a} where the stellar winds
from the B7 star BD+31~4224 have physically interacted with the remnant's
expanding shock front implies this star lies either inside or is in immediate
contact with the remnant's expanding shock front. This firmly anchors the
remnant to a distance close to that estimated for this star of
$726^{+13}_{-11}$. However, due to the star's half chord length of $\sim5.4$ pc
to the remnant center in the star's sight line along the remnant's northwestern
limb, this increases the Cygnus Loop's center distance uncertainty range by
roughly 5 pc to $726^{+18}_{-16}$ pc, corresponding to 710 thru 744 pc.

Since blue and red absorption \ion{Na}{1} lines from the Cygnus Loop's front
and rear expanding shells were observed in the spectra of both Stars X and Y,
their EDR3 estimated distances can be used to set additional limits on the
remnant's distance. Ignoring the Cygnus Loop's southern blowout region and
adopting a center near RA (J2000) = 20:51:11.2, Dec (J2000) = +31:03:10 for the
remnant's main emission structure, its 2.88${\degree}$ angular diameter implies
a physical radius of $\simeq18.5$ pc at the \citet{Fesen2018b} estimated
distance of 735 pc. Stars X and Y lie approximately 87\% and 84\%,
respectively, of the remnant's radius away from its center, meaning the half
chord lengths along their sight lines from the remnant's perpendicular center
line to its rear hemispheric edge are 9.3 pc and 10.2 pc, respectively. This
effectively decreases the possible location of the Cygnus Loop's center
distance using the distances to these background stars, namely $723\pm12$ pc
for Star X and  $698\pm11$ pc for Star Y.

The smaller distance to Star Y is especially restrictive since we know it has
to lie completely behind the remnant. Even assuming a tilt of the
remnant's eastern region toward us as described by \citet{Fesen2018b},
remnant center distances above $\simeq$740 pc seem unlikely given Star Y's
maximum estimated distance of 718 pc. We are thus left with a small range of
possible distances to the B7 star and hence the Cygnus Loop's center of 710 to
740 pc or $725\pm15$ pc. 

This new distance estimate can be combined with proper motions to determine the
speeds of radiative shock waves in the western Cygnus Loop, and those speeds
yield good agreement between models and observed spectra \citep{Raymond2020}.
However, the tight constraint on distance      exacerbates a problem.  Proper
motions of the northern nonradiative shocks and a distance of 725 pc imply
shock speeds that are a bit too small to match the thermal energies given by the
electron temperatures from X-rays and the proton temperatures from the
H$\alpha$ profiles \citep{Salvesen2009,Medina2014}.  New proper motion
measurements of the northern filaments \citep{Milanovic2019} generally agree
with the  proper motions of \citet{Salvesen2009} and the shock speeds derived
by \citet{Medina2014}.  Consequently, the problem might result either from
errors in proper motion measurements or from an inaccuracy in the models that
relate the H$\alpha$ line width to the proton temperature.

This slightly shorter distance estimate than the 735$\pm25$ pc of
\citet{Fesen2018b} does not appreciably alter any of the Cygnus Loop's
properties derived in that paper. But it does mean that the Cygnus Loop remnant
now has an estimated distance uncertainty comparable to that of its $\simeq$18 pc
radius, a remarkable result and a testament to the power of Gaia's accurate
parallax measurements. But given that this estimate depends heavily on the
assumption that the stellar winds from the B7 star BD+31~4224 have physically
interacted with the Cygnus Loop's expanding shocks thereby firmly anchoring the
remnant's distance, finding additional stars at varying distances toward the
Cygnus Loop showing high-velocity absorption lines in their spectrum would be
valuable in providing independent tests of this result.

\section*{Acknowledgements}

This work made use of data from the European Space Agency mission
{\it Gaia} (\url{https://www.cosmos.esa.int/gaia}), processed by the {\it Gaia}
Data Processing and Analysis Consortium (DPAC,
\url{https://www.cosmos.esa.int/web/gaia/dpac/consortium}). Funding for the DPAC
has been provided by national institutions, in particular the institutions
participating in the {\it Gaia} Multilateral Agreement.

\section*{Data Availability}

Parallax data are available through the
Gaia archive website.



\bibliographystyle{mnras}
\bibliography{ref2.bib} 



\bsp	
\label{lastpage}
\end{document}